\documentclass[twocolumn,showpacs,pre,aps]{revtex4}

\usepackage{dcolumn}
\usepackage{bm}
\usepackage{russian}

\begin{document}

\title{Fluctuation-Dissipation Relations for Continuous
Quantum Measurements}

\author{Yuriy E. Kuzovlev}
\email{kuzovlev@kinetic.ac.donetsk.ua} \affiliation{Donetsk
Physics and Technology Institute, 83114 Donetsk, Ukraine}


\begin{abstract}
The generating functional is derived for the
fluctuation-dissipation relations which result from the unitarity
and reversibility of microscopic dynamics and connect various
statistical characteristics of many consecutive (continuous)
observations in a quantum system subjected to external
perturbations. Consequences of these relations in respect to the
earlier suggested stochastic representation of interaction between
two systems are considered.
\end{abstract}


\maketitle

\section{Introduction}
Both the classical and quantum mechanics is time-reversible and
unitary (conserves classical phase volume and quantum
probability). In thermodynamic systems (ensembles), the unitarity
manifests itself in strong connections between noise and
dissipation, while the reversibility in time symmetry of noise and
reciprocity of transport processes. The famous examples are the
Einstein relation~\cite{ein}, the Nyquist formula~\cite{nyq}, and
the Onsager reciprocity relations~\cite{ons}. All these are the
relations between (i) second-order (quadratic) correlators of
equilibrium noise and (ii) linear parts of complete, possibly
nonlinear, responses to external perturbations (the expansion of
the responses in a series over powers of ``perturbing forces'' is
meant). Later, the fluctuation-dissipation theorem (FDT) proved by
Callen and Welton~\cite{cv} and the Green-Kubo formulas
~\cite{green,kubo} exhausted this linear theory.

The first nonlinear generalizations were obtained by
Efremov~\cite{efr} who proved the quadratic FDT, which connects
(i) quadratic parts of the responses and (ii) third-order (cubic)
equilibrium correlators ((iii) linear responses of quadratic
correlators also take a part here, but can be excluded). The next,
fourth-order, relations connect together (i) cubic components of
the responses, (ii) equilibrium fourth-order correlators and,
besides, (iii) quadratic responses of quadratic non-equilibrium
correlators ((iv) linear responses of cubic correlators are in
play too, but can be excluded). Their investigation was started by
Stratonovich~\cite{strat} who found that ``cubic FDT'' does not
exist (for details and more references see e.g.
\cite{strb,bk1,bk2}). Nevertheless, the fourth-order relations are
much useful, for instance, when analysing low-frequency
fluctuations in transport and relaxation rates, especially flicker
fluctuations \cite{diss,bk5,k1}.

The producing formulas for the whole (infinite) chain of
arbitrary-order fluctuation-dissipation and reciprocity relations
(FDR) were obtained in \cite{bk2}. In the works \cite{diss,bk3}
these results were extended to non-equilibrium steady states of
open systems. In \cite{bk1,diss} the extension to the thermic
perturbations was developed, that is perturbations of a
probability distribution (density matrix) of the system, in
addition to perturbations of its Hamiltonian which are termed
dynamic ones. Examples of various applications of FDR can be found
e.g. in \cite{bk1,diss,bk5,k1,bk3,bk4,bkt,bk6,k2}.

It is desirable to combine all the infinite variety of
arbitrary-order FDR into a compact visual generating FDR for the
probabilistic functionals or corresponding characteristic
functionals. In the framework of classical mechanics, this was
realized in \cite{bk2,bk3} (for review, see \cite{bk1}).

In the quantum theory, time-differed values of any interesting
variable $\,X(t)\,$ (an operator in the Heisenberg representation)
do not commute one with another, $\,X(t_1)X(t_2)\neq $
$X(t_2)X(t_1)$. But its measurements in macroscopic devices are
subjected to the classical description language, being thought as
a {\bf commutative} stochastic process, $\,x(t)\,$, whose values
are usual $c$-numbers. Hence, neither probabilistic nor
characteristic functional of $\,x(t)\,$ has a sense, until a
concrete definition of all $\,x(t)$'s correlators (statistical
moments), in terms of the $\,X(t)\,$, is chosen.

In general, the two lowest-order correlators only seem
unambiguously defined:
\begin{equation}
\begin{array}{c}
\left\langle x(t)\right\rangle
\equiv\,\text{Tr\,}X(t)\,\rho _0\,\,,\\
\left\langle x(t_1)x(t_2)\right\rangle
\equiv\,\text{Tr\,}(X(t_1)\circ X(t_2))\,\rho _0\,\,,\label{tf}
\end{array}
\end{equation}
with $\,\rho _0\,$ being the statistical operator (density
matrix), and $\,\,\circ\,$ designating the symmetric product
(Jordan product), $\,\,A\circ B\,\equiv\,(AB+BA)/2\,$. The
subscript ``0'' of $\,\rho _0$ means that $\,\rho _0$ represents a
quantum statistical ensemble at a fixed time moment, e.g. $\,t=0$.
The right-hand sides present definitions of the angle brackets on
the left sides, that is effective statistics of classical image,
$\,x(t)\,$, of the quantum variable $\,X(t)$.

What is for the higher-order correlators $\,\left\langle
x(t_1)x(t_2)...x(t_N) \right\rangle\,$, unfortunately, at any
$\,N>2\,$, there are $\,N!/2>1\,$ different symmetrized
(Hermitian) expressions produced by various permutations of
$\,X(t_k)\,$ in $\frac
12\,$Tr$\,(X(t_1)X(t_2)...X(t_N)+X(t_N)...X(t_2)X(t_1))\rho _0\,$,
plus uncountably many their weighted linear combinations.

Moreover, in fact even the second correlator can be introduced in
an alternative way. For example, let us define the characteristic
functional (CF) of $\,x(t)\,$ by the identity
\begin{equation}
\begin{array}{c}
\left\langle \,\exp\left[\int_0^t
v(t^{\prime})x(t^{\prime})dt^{\prime}\right]\,\right\rangle\equiv
\\
\equiv\text{Tr\,}\exp\left[\int_0^t
v(t^{\prime})X(t^{\prime})dt^{\prime}+\,\ln\,\rho _0\right]
\,\,\label{pf0}\\
\end{array}
\end{equation}
Here $\,t>0$, and $\,v(t)\,$ is an ``arbitrary probe function''
(test function). Double differentiation of (\ref{pf0}) by
$\,v(t_1)$ and $\,v(t_2)$ at $\,v(t)\equiv 0$ gives
\begin{equation}
\left\langle x(t_1)x(t_2)\right\rangle= \text{Tr\,}\int_0^1
X(t_1)\rho_0^{\alpha}X(t_2)\rho_0^{1-\alpha}d\alpha
\,\,,\label{pf1}\\
\end{equation}
and $N$-order moments include $\,N$ different $\,\rho_0$'s powers
whose sum equals to unit. As pointed out in \cite{bk1}, under this
specific definition of the CF all the FDR between the angle
brackets look absolutely similar in both classical and quantum
case.

Of course, to become practically useful, a choice of definition of
the CF, i.e. definition of all statistical moments $\,\langle
x(t_1)$ $...$ $x(t_N)\rangle$, should be based on analysis of real
measurement procedures. The well known examples presents quantum
theory of electromagnetic field fluctuations (see
e.g.~\cite{lax}). During last decade, in the original works by
Levitov, Lesovik, Nazarov and others~\cite{ll1,ll2,naz,les} (see
also references therein) the continuous measurements of the charge
transport in electric devices were analyzed. The results of these
works allow to conclude that frequently the adequate construction
rule of the CF is the chronological symmetrized product:
\begin{equation}
\label{pf}
\begin{array}{c}
\left\langle \,\exp\left(\int_0^t
v(t^{\prime})x(t^{\prime})dt^{\prime}\right)\,\right\rangle \equiv
\\
\text{Tr\,}\,\overleftarrow{\exp}\left(\frac 12 \int_0^t
v(t^{\prime})X(t^{\prime})dt^{\prime}\right)\rho _0\,\,
\overrightarrow{\exp}\left(\frac 12 \int_0^t
v(t^{\prime})X(t^{\prime})dt^{\prime}\right)
\end{array}
\end{equation}
Here $\,\overleftarrow{\exp}$ and $\overrightarrow{\exp}$ are
chronological and anti-chronological exponents, respectively.
According to this rule (and to general properties of the trace
operation ~{\bf Tr}), the quadratic correlator remains as in
(\ref{tf}). For the higher-order correlators the rule (\ref{pf})
prescribes
\begin{equation}
\begin{array}{c}
\langle x(t_1)x(t_2)x(t_3)\rangle = \text{Tr\,} (X(t_1)\circ (
X(t_2)\circ X(t_3)))\,\rho _0\,\,, \\
\langle x(t_1)x(t_2)x(t_3)x(t_4)\rangle =\, \\
\,=\text{Tr\,} (X(t_1)\circ ( X(t_2)\circ (X(t_3)\circ
X(t_4))))\,\rho _0\,\,,
\end{array} \label{34}
\end{equation}
and so on, where, for definiteness, the inequalities $\,\,t_N\geq
...$ $t_2\geq t_1\,\,$ are presumed.

Below, we will derive generating FDR for so built statistical
moments. All the more this is interesting because the same rule
(\ref{pf}) naturally arose in the course of the so-called
``stochastic representation of deterministic interactions''
~\cite{ya1,ya2,ya3,ya4} (the general method for correct
construction of ``Langevin equations'' introducing thermodynamic
noise and dissipation into quantum or classical dynamics).

The strong argument for the benefit of the rule (\ref{pf}) is that
it can be deduced from the correspondence principle. To see this,
firstly consider classical systems.

\section{Classical characteristic functionals}
Let a classical system has canonical variables (coordinates and
momenta) $\,\Gamma=$ $\{q,$ $p\}$. Generally, the system undergoes
a dynamic perturbation from its outside, which means that its
Hamiltonian, $\,H_t=$ $H_t(\Gamma)\,$, is time dependent.
Introduce also the corresponding Liouville operator $\,L_t\,$, and
the evolution operator $\,Z_t\,$:
\[
L_t=(\nabla _q H_t)\nabla _p-(\nabla _p H_t)\nabla _q\,\,,\,\,\,\,
Z_t= \overleftarrow {\exp}\left[\int _0^tL_{\tau }d\tau \right]\,
\]
Eventually, we are interested in the evolution and fluctuations of
variables $\,X\,$ which represent definite functions of the phase
space point currently occupied by the system: $\,X=$ $X(\Gamma
)\,$ (breafly, ``phase functions'').

The essential properties of the Liouville and evolution operators
are as follow:
\begin{equation}
\begin{array}{c}
\int\,A(\Gamma )L_t B(\Gamma )\,d\Gamma=-\int\,B(\Gamma )L_t
A(\Gamma )\,d\Gamma\,\,\,,\\
\int\,Z_t\rho(\Gamma )\,d\Gamma=\int\,\rho(\Gamma)\,d\Gamma\,\,,\\
Z_t^{-1}\Gamma=\Gamma _t(\Gamma )\,\,\,,\\ \label{props}
Z_t^{-1}X(\Gamma )Z_t=X(Z_t^{-1}\Gamma )\,\,,
\end{array}
\end{equation}
where $\,A(\Gamma )$, $B(\Gamma )$, $X(\Gamma )$ and $\rho(\Gamma
)\,$ are any phase functions (such that $\,A(\Gamma )B(\Gamma )$
and $\rho(\Gamma )\,$ are integrable), and $\,\Gamma _t(\Gamma
)\,$ stand for the current values of the canonic variables (at
time $\,t\,$) expressed through their initial values $\,\Gamma $
(at the initial time moment $\,t=0$). In other words, $\,\Gamma
_t(\Gamma )\,$ is the solution of the Hamilton equations under
initial condition $\,\Gamma _0(\Gamma )$ $=\Gamma $.

Importantly, the latter equality in (\ref{props}) is operator
equality, that is both $\,X(\Gamma )\,$ on the left and
$\,X(Z_t^{-1}\Gamma )\,$ on the right have the sense of
multiplication operators. Combination of this equality with the
previous one implies
\begin{equation}
Z_t^{-1}X(\Gamma )Z_t=X(t,\Gamma )\equiv X(\Gamma _t(\Gamma ))\,
\label{prop}
\end{equation}
Hence, the composite operator $\,Z_t^{-1}X(\Gamma )Z_t$ reduces to
operator of multiplication by the time-dependent number
$\,X(t,\Gamma )\,$, which is nothing but trajectory of the
variable $\,X\,$ under initial conditions $\,\Gamma $.

Next, let us be convinced that CF of any variable (phase function)
$\,X(t,\Gamma )$ $\equiv X(\Gamma _t(\Gamma))\,$ can be expressed
\cite{bk1} by the formula
\begin{equation}
\begin{array}{c}
\left\langle \exp \left[\int _0^t v(\tau )x(\tau ) d\tau
\right]\right\rangle =   \label{ccf}\\
\,\,\,=\int \overleftarrow {\exp} \left(\int _0^t [L_{\tau }+
v(\tau)X(\Gamma )]\,d\tau \right)\,\rho _0(\Gamma) d\Gamma
\end{array}
\end{equation}
Here $\,\rho _0\,$ is the statistical operator, i.e. distribution
function, of the system at time $\,t=0$, and again $\,v(t)\,$ is
arbitrary probe function. In the angle brackets, $\,x(t)\,$ means
the internal variable $\,X(t,\Gamma)$ as perceived by an outside
observer (which knows nothing about $\,\Gamma $) and interpreted
by him as a stochastic process.

To justify (\ref{ccf}), it is sufficient to make standard
``disentangling'' of the complex exponent in the lower row of
(\ref{ccf}):
\begin{equation}
\begin{array}{c}
\int\overleftarrow {\exp} \left[\int _0^t [L_{\tau }+
v(\tau)X(\Gamma )]\,d\tau \right]\,\rho_0(\Gamma)d\Gamma=\\
=\int Z_t\,\overleftarrow {\exp} \left[\int _0^t v(\tau)Z_{\tau
}^{-1}X(\Gamma )Z_{\tau }\,d\tau \right]\,
\rho_0(\Gamma)d\Gamma=\\
=\int\,\overleftarrow {\exp} \left[\int _0^t v(\tau)X(\tau,\Gamma)
\,d\tau \right]\,\rho_0(\Gamma)d\Gamma\,\label{dis}
\end{array}
\end{equation}
Here, the latter transformation is made with taking into account
the identity (\ref{prop}) and the second property from
(\ref{props}). Evidently, final expression in (\ref{dis})
coincides with what is meant in the angle brackets in (\ref{ccf}),
i.e. the CF of the path $\,X(t,\Gamma)\,$ whose uncontrolled
dependence on the initial conditions $\,\Gamma\,$ turns it into a
random process, $\,x(t)\,$.

According to (\ref{ccf})-(\ref{dis}), evaluation of the CF is
equivalent to solution of definite differential equation:
\begin{equation}
\begin{array}{c}
\left\langle \exp \int _0^t v(\tau )x(\tau ) d\tau
\right\rangle =\int \rho\,d\Gamma\,\,\,,\\
d\rho /dt\,=\,[L_t+v(t)X(\Gamma)]\rho\,\,\,,\label{de}
\end{array}
\end{equation}
where the function $\,\rho =\rho(t,\Gamma)\,$ satisfies the
initial condition $\rho(0,\Gamma)$ $=\rho _0(\Gamma)$.

In principle, all what just was said is known. The representation
(\ref{de}), or (\ref{ccf}), used in~\cite{bk1,bk3}, is variation
of so-called Feynman-Kac formulas~\cite{rs} which connect path
integrals and differential equations. In fact, this representation
of the CF is valid not for deterministic evolution only, but also
for Marcovian stochastic evolutions. In this case, $\,\Gamma$
designates instant state of a (multi-component) Marcovian random
process, and $\,L_t\,$ its evolution (kinetic) operator
\cite{bk1,diss,bk3}.

\section{Quantum characteristic functionals}
Following the correspondence principle, it seems natural to
suggest formulas (\ref{ccf}) and/or (\ref{de}) be the basis for
definition of the CF in quantum case.

Now, the Liouville operator changes to the commutator: $\,L_t\Phi
=$ $i[\Phi ,$ $H_t]/\hbar\,$ (with $\,[A,B]\equiv $ $AB-BA\,$, and
$\,\Phi ,A,B\,$ being arbitrary operators). Quantum analogue of
the $\,Z_t$ exploited in previous section is super-operator whose
action is defined by
\[
\begin{array}{c}
Z_t\,\rho =U(t)\rho U^{-1}(t)\,\,,\,\,\,U(t)\equiv\overleftarrow
{\exp}[-\frac {i}{\hbar }\int _0^tH_{\tau }d\tau ]
\end{array}
\]
The phase functions $\,X(\Gamma )\,$ and $\,X(\Gamma _t(\Gamma
))\,$ are replaced by the operators of quantum variable
(observable) in the Shrodinger and Heisenberg representation,
respectively, $\,X\,$ and $\,X(t)$, where $\,X(t)=
\,Z_t^{-1}X\,=\,U^{-1}(t)XU(t)\,$.

What is for the operation of multiplication by $\,X(\Gamma )\,$ in
(\ref{de}), it should be replaced by super-operator of the
symmetric product: $\,X(\Gamma )\Phi (\Gamma )$ $\Rightarrow$
$X\circ \Phi\,$. The matter is that it is very hard to suggest
something else. Under this extension, the equality (\ref{prop})
remains valid:
\[
\begin{array}{c}
Z_t^{-1}(X\circ(Z_t\Phi))=X(t)\circ\Phi
=(U^{-1}(t)XU(t))\circ\Phi\,
\end{array}
\]
(but now, generally, the super-operator $\,X(t)\circ\,$ in none
base reduces to scalar multiplication).

Thus one comes to the construction of quantum CF as follows:
\begin{equation}
\left\langle \exp \left[\int _0^t v(\tau )x(\tau ) d\tau
\right]\right\rangle =\,\text{Tr\,} \rho\,\,\,,\label{rho}
\end{equation}
\begin{equation}
\frac {d\rho}{dt}=\frac {i}{\hbar}[\rho
,H_t]+v(t)X\circ\rho\,\,\,,\label{qde}
\end{equation}
with the initial condition $\rho(t=0)$ $=\rho _0$. One can easy
verify that substitution of the formal direct solution of
(\ref{qde}) into (\ref{rho}) produces just the formula (\ref{pf}),
with $\,X(t)\,$ standing for Heisenbergian operator of the quantum
variable and $\,x(t)\,$ its effective commutative image.

The doubtless advantage of such definition,
(\ref{rho})-(\ref{qde}), of quantum CF is its differential
nature~\cite{ya1,ya2}, which highlights its automatic agreement
with the causality principle too. The formulas (\ref{34})
demonstrate that because of the causality all the corresponding
higher-order statistical moments appear asymmetric with respect to
time inversion. It is natural: earlier observations (measurements)
can affect later ones, but opposite influence is impossible.

Further, we want to extend (\ref{pf}) to an arbitrary set of
variables under simultaneous continuous observation. With this
purpose, let us replace $\,v(t)X\,$ in (\ref{qde}) with the
operator
\begin{equation}
V_t=\sum_{\mu \nu } v_t^{\mu \nu }X_{\mu \nu }\,\,,\,\,\, X_{\mu
\nu }\equiv |\mu \rangle\langle \nu |\,\,,    \label{vx}
\end{equation}
where $\,|\mu\rangle\,$ are states, or vectors (in the Dirac's
designations), which constitute a complete orthonormal base, and
$\,v_t^{\mu \nu }\,$ arbitrary probe functions. This is most
general form of the observation. Analogously, most general
external perturbation can be described as
\begin{equation}
H_t=H_0-F_t\,\,,\,\,\,F_t= \sum_{\mu \nu } f_t^{\mu \nu }X_{\mu
\nu}\,\,, \label{fx}
\end{equation}
where $\,H_0$ is Hamiltonian of the ``free'' system, and
$\,f_t^{\mu \nu }\,$ are ``perturbing forces''.

Then, instead of (\ref{qde}) and (\ref{pf}), we have
\begin{equation}
\frac {d\rho}{dt}=\frac {i}{\hbar}[\rho
,H_0-F_t]+V_t\circ\rho\,\,\,,\label{ee}
\end{equation}
\begin{equation}
\begin{array}{c}
\left\langle \exp\left[\int _0^t v_{\tau }^{\mu \nu }x_{\mu \nu
}(\tau ) d\tau \right]\right\rangle _{H_0,F_{\tau}}=\,\text{Tr\,}
\rho\,\equiv \Xi (V_{\tau };\,H_0,F_{\tau }) \label{qf0}
\end{array}
\end{equation}
From here, the repeated indices $\,\mu\,$ or $\,\nu\,$ mean
summation over them; ~$\,x_{\mu \nu }(t)\,$ are effective
classical images of quantum variables $\,X_{\mu \nu
}(t)=U^{-1}(t)X_{\mu \nu }U(t)\,$; ~$\,\Xi (V_{\tau };H_0,F_{\tau
})\,$ will be used as shortened designation of the CF, i.e. the
angle brackets.

The second and third arguments of $\,\Xi$, as well as the
subscript under angle brackets, remind about eigen Hamiltonian of
the system and its perturbations. Direct solution of (\ref{ee})
yields
\begin{equation}
\begin{array}{c}
\Xi (V_{\tau };H_0,\,F_{\tau })=\\
=\text{Tr\,}\,\,\overleftarrow{\exp } \left(\int _0^t\left[ -\frac
i\hbar (H_0-F_{\tau })+\frac 12 V_{\tau }\right] d\tau
\right)\,\rho _0\times \\
\,\,\,\,\,\,\,\,\,\,\times \,\overrightarrow{\exp } \left(\int
_0^t\left[ \frac i\hbar (H_0-F_{\tau })+\frac 12 V_{\tau
}\right] d\tau \right)= \\
=\text{Tr\,}\,\,\overleftarrow{\exp } \left[\frac 12\int
_0^tv_{\tau}^{\mu \nu }X_{\mu \nu }(\tau) d\tau
\right]\,\rho _0\times \\
\,\,\,\,\,\,\,\,\,\,\,\,\times \,\overrightarrow{\exp }
\left[\frac 12\int _0^tv_{\tau}^{\mu \nu }X_{\mu \nu }(\tau) d\tau
\right]\,
\end{array} \label{qf}
\end{equation}
The $\,\Xi$'s arguments $\,V_t$ and $\,F_t\,$ can be understood as
either two sets, of the probe functions $\,v_t^{\mu \nu }\,$ and
the forces $\,f_t^{\mu \nu }\,$, or the whole operators introduced
in (\ref{vx}) and (\ref{fx}). Of course, the pair $\,x_{\mu \nu
}(t)\,$ and $\,x_{\nu \mu }(t)\,$ in (\ref{qf0}) can be
interpreted as two mutually conjugated complex-valued random
processes (at least while $\,F_t\,$ assumed Hermitian).

\section{The generating FDR}
In the present paper we confine ourselves by the assumption that
the initial density matrix, $\,\rho_0\,$, is canonic
thermodynamically equilibrium one: $\,\rho_0\propto \exp
(-H_0/T)\,$ (in the classical theory, not only this case already
was considered~\cite{bk2,bk3} but also the case of non-equilibrium
initial distributions~\cite{bk1,diss}).

Following the recipes of~\cite{bk2}, let us make three
transformations of the operator expression placed after the trace
symbol in (\ref{qf}).

(i) Firstly, transpose this expression as a whole, with the help
of the usual rule $\,(AB...C)^{\prime}=$
$C^{\prime}...B^{\prime}A^{\prime}\,$, where the prime means the
transposition: $\,A^{\prime}=$ $(A^{\dag})^{\ast}=$
$(A^{\ast})^{\dag}\,$ (symbols $\,\dag\,$ and $\,\ast\,$ will
stand for the Hermitian and complex conjugation, respectively). As
is well known, this operation does not change the trace.

(ii) Secondly, invert the direction of the time account, by means
of rewriting chronological exponents as anti-chronological ones,
and vice versa, as in the examples
\[
\begin{array}{c}
\overleftarrow{\overrightarrow{\exp }} \left
[\int_0^tA(\tau)d\tau\right ]= \overrightarrow{\overleftarrow{\exp
}} \left [\int _0^t A(t-\tau)d\tau\right ]
\end{array}
\]

(iii) Thirdly, rewrite $\,\rho _0\,$ in the form $\,\sqrt{\rho
_0}$$\sqrt{\rho _0}\,$ and ``drag'' one of these multiplicands to
the left and another to the right and after that again unify them,
with the help of the trace property ~Tr$\,ABC\,=$~Tr$\,CAB\,$.

The result of these manipulations is an expression which looks
quite identical to the initial one, but with modified operator of
the observation instead of $\,V_t\,$ and modified operator of the
perturbation instead of $\,F_t\,$. To write the result, of course,
it is convenient to choose the states $\,|\mu\rangle\,$ as a
complete set of orthonormal eigenvectors of the free Hamiltonian
$\,H_0\,$: $\,H_0|\mu\rangle=E_{\mu}|\mu\rangle$, with the
eigenvalues (energies) $\,E_{\mu}$. Besides, introduce two
operators
\[
\begin{array}{c}
L_0\Phi\equiv i[\Phi ,H_0]/\hbar\,\,,\,\,\,U_0(t)\equiv\exp[-iH_0
t/\hbar]\,\,,
\end{array}
\]
the matrices (in terms of the chosen basis)
\begin{equation}
S_{\nu \mu }\,,C_{\nu \mu }\equiv \sinh\,,\cosh \frac {E_{\nu \mu
}}{2T}\,\,,\,\,\,
 \Delta _{\mu \nu }\equiv
\frac {2T}{E_{\mu \nu }} \tanh \frac { E_{\mu \nu
}}{2T}\,\,,\label{coefs}
\end{equation}
and super-operators $\,\bm{C}$, $\,\bm{S}\,$ and $\,\bm{\Delta}\,$
whose action is determined by these matrices:
\begin{equation}
\begin{array}{c}
\bm{C}=\cos \left(\frac {\hbar}{2T}L_0\right)\,\,,\,\,\, (\bm{C}
\Phi )_{\mu \nu }=C_{\mu \nu }\Phi _{\mu \nu }\,\,,
\,\,\,\,\,\,\,\,\\
\,\,\,\,\,\,\,\,\bm{S}=\sin \left(\frac
{\hbar}{2T}L_0\right)\,\,,\,\,\,
(\bm{S} \Phi )_{\mu \nu }=S_{\mu \nu }\Phi _{\mu \nu }\,\,,\\
\,\,\,\,\,\,\,\,\,\,\,\,(\bm{\Delta}\Phi )_{\mu\nu
}=\Delta_{\mu\nu }\Phi _{\mu\nu }
\end{array}   \label{scd}
\end{equation}

Then, finally, we can formulate the generating FDR. It is
expressed by the formulas as follow:
\begin{equation}
\,\,\,\,\,\,\Xi (V_{t-\tau }^{\prime}\,;\,
H_0^{\prime},\,F_{t-\tau }^{\prime})= \Xi (\widetilde {V}_{\tau
}\,;\,H_0,\, \widetilde {F}_{\tau })\,\,, \label{fds}
\end{equation}
where $\,V_{\tau}^{\prime}=v_{\tau }^{\mu \nu}X_{\mu
\nu}^{\prime}\,$, $\,\widetilde {V}_{\tau}=\widetilde{v}_{\tau
}^{\mu \nu}X_{\mu\nu}\,$, $\,F_{\tau}^{\prime}=f_{\tau }^{\mu
\nu}X_{\mu \nu}^{\prime}\,$, $\,\widetilde
{F}_{\tau}=\widetilde{f}_{\tau }^{\mu \nu}X_{\mu\nu}\,$, and
\begin{equation}
\left[
\begin{array}{c}
\widetilde{f}_{\tau }^{\mu \nu} \\
\widetilde{v}_{\tau }^{\mu \nu}
\end{array}
\right]= \left[
\begin{array}{cc}
C_{\mu \nu } & \frac {i\hbar }{2} S_{\mu \nu } \\
\frac {2}{i\hbar }S_{\mu \nu } & C_{\mu \nu }
\end{array} \right] \left[
\begin{array}{c}
f_{\tau }^{\mu \nu} \\ v_{\tau }^{\mu \nu}
\end{array} \right]\,\,\,,      \label{nv2}
\end{equation}
or, in another equivalent form,
\begin{equation}
\left[
\begin{array}{c}
\widetilde{F}_{\tau } \\ \widetilde{V}_{\tau }
\end{array}
\right] \equiv \left[
\begin{array}{cc}
\bm{C} &
-\frac{\hbar }{2}\bm{S} \\
\frac{2}{\hbar }\bm{S} & \bm{C}
\end{array} \right] \left[
\begin{array}{c}
F_{\tau } \\ V_{\tau }
\end{array} \right]              \label{nv1}
\end{equation}
For convenience, the observation and perturbation variables are
unified into the column vector.

The combined transformation
$\,\Phi_{\tau}\Leftrightarrow\Phi_{t-\tau}^{\prime}\,$ represents
the time reversal. Thus, (\ref{fds}) and (\ref{nv2}) or
(\ref{nv1}) state invariance of the CF under (i) simultaneous time
reversal of both the probe functions and external forces and (ii)
their mutual mixing as described by (\ref{nv2}) and (\ref{nv1}).

Importantly, when dealing with (\ref{fds})-(\ref{nv2}) one should
remember about the time translational invariance: any joint
temporal shift of $\,V_t\,$ and $\,F_t\,$ does not change
$\,\Xi(V_{\tau };H_0,F_{\tau })$'s value, since $\,\rho _0\,$ is
invariant with respect to free (unperturbed and unobserved)
evolution ($\,[\rho_0,H_0]=0\,$).

It should be underlined also that by the very definition (see
(\ref{vx})) of the operators $\,X_{\mu \nu }\,$ we have
\begin{equation}
X^{\prime}_{\mu\nu }=|\nu^{\ast}\rangle\langle\mu^{\ast}|\,\,,
\,\,\,\,\,H_0^{\prime}|\mu^{\ast}\rangle
=E_{\mu}|\mu^{\ast}\rangle\,\,, \label{xprime}
\end{equation}
where $\,|\mu^{\ast}\rangle\,$ are eigenvectors of the transposed
free Hamiltonian $\,H_0^{\prime}\,$, with the same eigenvalues
$\,E_{\mu}\,$ as that of $\,|\mu\rangle\,$ (one can write also
$\,|\mu^{\ast}\rangle=$ $|\mu\rangle^{\ast}=$
$\langle\mu|^{\prime}\,$). Hence, during the time reversed
evolution the operator $\,X^{\prime}_{\mu\nu }\,$ represents those
quantum variable which is represented by operator $\,X_{\nu\mu
}\,$ in the direct process. Consequently, in terms of their
effective classical images, the left side of (\ref{fds}) looks as
\begin{equation}
\begin{array}{c}
\Xi (V_{t-\tau }^{\prime};\,H_0^{\prime},F_{t-\tau
}^{\prime})=\,\,\,\,\,\\\,\,\,\,\,=\left\langle \exp\left[\int
_0^t v_{t-\tau }^{\nu \mu }x_{\mu \nu }(\tau ) d\tau
\right]\right\rangle _{H_0^{\prime},\,F_{t-\tau}^{\prime}}\,
\label{ti}
\end{array}
\end{equation}
In other words, under the time reversal the matrix $\,\{v_{\tau
}^{\mu \nu }\}\,$ behaves like the operator $\,V_{\tau}\,$, that
is it undergoes transposition: $\,v_{\tau }^{\mu \nu
}\Leftrightarrow $ $(v_{t-\tau }^{\prime})^{\mu \nu }=$ $v_{t-\tau
}^{\nu \mu }\,$ (similarly, $\,f_{\tau }^{\mu \nu }\Leftrightarrow
$ $(f_{t-\tau }^{\prime})^{\mu \nu }=$ $f_{t-\tau }^{\nu \mu }$).

In the classical limit (formally $\,\hbar\rightarrow 0$), we have
 $\bm{C}\rightarrow 1$, $\,\bm{S}\rightarrow 0\,$,
$\,2\bm{S}/\hbar\rightarrow $ $T^{-1}L_0\,$, and formula
(\ref{nv1}) takes the form
\begin{equation}
\widetilde{F}_{\tau }=F_{\tau }\,\,,\,\,\,\widetilde{V}_{\tau
}=V_{\tau }+T^{-1}L_0F_{\tau }\,      \label{cl}
\end{equation}
At that, the operators $\,F_{\tau }$ and $\,V_{\tau }\,$ turn into
phase functions, $\,L_0\Rightarrow(\nabla _q H_0)\nabla
_p$$-(\nabla _p H_0)\nabla _q\,$, and the prime in (\ref{fds})
means replacement of $\,\{q,p\}\,$ with $\,\{q,-p\}$.

Comparison between (\ref{nv1}) and (\ref{cl}) reveals that the
observations anyway are influenced by the perturbations, but
opposite effects exist in the quantum theory only. For
illustration, let us choose $\,F_{\tau }\equiv 0\,$, that is the
real direct-time observations (described by $\,V_{\tau }$) are
made in equilibrium system. Then, according to
(\ref{fds})-(\ref{nv1}), detection of exactly time-reversed
results (described by $\,V_{t-\tau }^{\prime}\,$), in equilibrium
system too, is in certain sense the same as detection of the
direct results but under specific nonzero perturbations. This
unpleasant peculiarity of the quantum theory is not surprising:
since any measurement influences subsequent ones, but not vice
versa, the mere rearrangement of their results, generally
speaking, could not realize under same conditions.

Various particular FDR can be obtained from
(\ref{fds})-(\ref{nv1}) by either some special choice of
$\,V_{\tau }\,$ and $\,F_{\tau }\,$ or variational
differentiations with respect to $\,v_{\tau }^{\mu \nu }\,$ and
$\,f_{\tau }^{\mu \nu }$. Evidently, if both $\,F_{\tau }\,$ and
$\,V_{\tau }\,$ are Hermitian, then both transforms
$\,\widetilde{V}_{\tau }\,$ and $\,\widetilde{F}_{\tau }\,$ also
are Hermitian. But if $\,V_{\tau }\,$ is quite arbitrary, then one
should be ready to deal with non-Hermitian perturbation
$\,\widetilde{F}_{\tau }$.

For example, let us choose $\,V_t\equiv 0$, i.e. there is no
observation at all. According to (\ref{qde}) or (\ref{qf}), of
course, in absent of observations the evolution of the statistical
operator $\,\rho \,$ is unitary under whatever perturbations.
Therefore ~Tr$\,\rho\,=$~Tr$\rho_0=1\,$, and $\,\Xi(0;
H_0,\,F_{\tau })\equiv 1\,$. Consequently, (\ref{fds}) and
(\ref{nv1}), together with (\ref{scd}), yield the equality
\begin{equation}
1=\Xi \left(\frac {1}{T}\bm{\Delta}L_0\widetilde {F}_{\tau
}\,;\,H_0,\, \widetilde {F}_{\tau} \right)\,\,,\label{v0}
\end{equation}
or, in other equivalent designations,
\begin{equation}
\left\langle \exp\left[-\int _0^t \frac {2i}{\hbar} \tanh
\left(\frac {E_{\mu \nu}}{2T}\right)\widetilde{f}_{\tau}^{\mu\nu}
x_{\mu\nu}(\tau) d\tau \right]\right\rangle_{H_0,\,\widetilde
F_{\tau}}=1\, \label{v1}
\end{equation}
These relations are valid at {\bf arbitrary} perturbation operator
$\,\widetilde{F}_{\tau }\,$. The subscript under the angle bracket
reminds about the perturbation.

To make Eqs.\ref{v0}-\ref{v1} better transparent, combine them
with the identities
\[
\begin{array}{c}
L_0\widetilde {F}_t=\frac i{\hbar}[\widetilde {F}_t,H_0]=-\frac
i{\hbar}[H_0-\widetilde {F}_t,H_0]=-P_t
\end{array}
\]
where $\,P_t\,$ is operator of the instant energy power being
pumped by the perturbations (notice that $\,dH_0(t)/dt=$
$d(U^{-1}(t)H_0U(t))/dt$ $=U^{-1}(t)P_tU(t)\,$). Hence, we can
rewrite Eq.\ref{v0} also as
\begin{equation}
\begin{array}{c}
\Xi \left(-\frac {1}{T}\bm{\Delta}P_{\tau }\,;\,H_0,\, \widetilde
{F}_{\tau} \right)\,=1\,\,\,,
\end{array}      \label{pv}
\end{equation}
with arbitrary $\,\widetilde{F}_{\tau }\,$ and related
$\,P_t=\frac i{\hbar}[H_0,\widetilde {F}_t]$.

In the classical limit, $\,\bm{\Delta}\,$ disappears, in the sense
that $\,\Delta_{\mu\nu}\rightarrow1$. Therefore, (\ref{pv}) turns
into equality $\,\langle\exp(-A/T)\rangle=1\,$ \cite{bk1,bk2},
where $\,A\,$ is total work produced by perturbations during time
interval $\,(0,t)\,$. More generally, perhaps,
$\,\bm{\Delta}P_t/T\,$ can be interpreted as operator of entropy
production.

\section{Probability functionals}
Provided statistical moments of quantum variables $\,X(t)\,$ and
their CF are defined (by (\ref{ee})-(\ref{qf}), in our case), we
can introduce the probability functional (PF) of their classical
(commutative) equivalents $\,x(t)\,$. We will designate it by
$\,W(x(\tau )\,;\,H_0,\,F_{\tau })\,$. As usually, is represents
functional Fourier transform of the CF:
\begin{equation}
\begin{array}{c}
W(x(\tau )\,;\,H_0,\,F_{\tau })= \int \exp \left[-i\int_0^t
v_{\tau }^{\mu\nu}x_{\mu\nu}(\tau )d\tau \right]\times\\
\,\,\,\,\times \,\Xi (\,iV_{\tau }\,;\,H_0,\,F_{\tau
})\,dV_{\tau}\,\,\,,
\end{array}    \label{fur}
\end{equation}
where $\,dV_{\tau}= \prod
_{\tau\mu\nu}(dv_{\tau}^{\mu\nu}/2\pi)\,$.

Let us apply this transform to both sides of the FDR~(\ref{fds}),
again with the indices being related to the eigenstates of
$\,H_0$. We omit rather bulky manipulations and write their result
in two steps:
\begin{equation}
W(x^{\prime}(t-\tau )\,;\,H_0^{\prime},\, F_{t-\tau }^{\prime})=
\label{fur1}
\end{equation}
\[
=\exp\left[-\int _0^t \frac {2i}{\hbar} \tanh \left(\frac {E_{\mu
\nu}}{2T}\right)f_{\tau}^{\mu\nu} x_{\mu\nu}(\tau) d\tau
\right]\times
\]
\[ \times\,\widetilde{W}(x(\tau )\,;\,H_0,\,
F_{\tau })\,\,,
\]
\begin{equation}
\widetilde{W}(x(\tau )\,;\,H_0,\, F_{\tau })=  \label{fur2}
\end{equation}
\[=\exp\left[\int _0^t d\tau \frac
{\hbar}{2i}\,S_{\mu\nu}C_{\mu\nu}\,\frac {\delta}{\delta
f_{\tau}^{\mu\nu}} \frac {\delta}{\delta x_{\mu\nu}(\tau
)}\right]\times
\]
\[
\times\,W(\bm{C}^{-1}x(\tau )\,;\,H_0,\,\bm{C}^{-1}F_{\tau })\,\,,
\]
where $\,(\bm{C}^{-1}\Phi)_{\mu\nu}=\Phi_{\mu\nu}/C_{\mu\nu}$.

In these two formulas, both the exponents in middle rows result
from the left bottom and right top non-diagonal elements of matrix
(\ref{nv2}) (or (\ref{nv1})), respectively, i.e. from mutual
mixing of perturbation and observation. Thus, the exponent in
(\ref{fur2}) reflects disturbing action of observations, which is
quite unpleasant peculiarity of the quantum case. Of course, in
fact this operator-valued exponent acts as an integral operator.
We leave it unwrapped till possible separate work (though example
of similar operators can be found in \cite{ya4}). For the present,
confine ourselves by the sad conclusion that generally
$\,\widetilde{W}(x(\tau );H_0,F_{\tau })\,$, and thus PF of the
reversed process, $\,W(x^{\prime}(t-\tau );H_0^{\prime},F_{t-\tau
}^{\prime})\,$, relates to PF of the direct process $\,W(x(\tau
);H_0,F_{\tau })\,$ in some {\bf non-local} way, with respect to
both $\,f_t^{\mu\nu}\,$ and $\,x_{\mu\nu}(t)$.

However, it is not hard to notice that a degree of the
non-locality is proportional to $\,\tanh ^2(E_{\mu\nu}/2T)\,$,
hence, it is negligible in respect to low-energy quantum
transitions. In the classical limit ($\hbar\rightarrow 0$) the
exponent in (\ref{fur2}) turns into unit,
$\,\widetilde{W}\rightarrow W\,$, and (\ref{fur1})-(\ref{fur2})
reduce to the purely {\bf local} relation
\begin{equation}
\begin{array}{c}
W(x^{\prime}(t-\tau );H_0^{\prime},F_{t-\tau }^{\prime})=
\exp(-A/T)\,W(x(\tau );H_0,F_{\tau })\,\,,\\
\,\,\,\,\,\,\,A=\int _0^t f_{\tau}^{\mu\nu}\frac
{d}{d\tau}x_{\mu\nu}(\tau)\,d\tau\,\,\,,
\end{array}   \label{clpf}
\end{equation}
where $\,A\,$ is again the work of the external forces
$\,f_t^{\mu\nu}\,$, and $\,\Phi^{\prime}(q,p)\equiv \Phi(q,-p)$.
This FDR is equivalent to what was obtained in \cite{bk2}.
Curiously, if quantum CF is defined by Eq.\ref{pf0}, instead of
Eqs.\ref{ee}-\ref{qf}, then Eq.\ref{clpf} replaces
Eqs.\ref{fur1}-\ref{fur2} even in general quantum case.

\section{Stochastic representation of $\,\,\,\,\,\,\,\,\,$
response to perturbations}
The above consideration demonstrated a
lot of formal symmetry between perturbations and observations.
This symmetry suggests that the perturbing forces
$\,f_t^{\mu\nu}\,$ can be treated as another test (probe)
functions which correspond to watching for additional ghost
variables. Let the latter be named as $\,y_{\mu\nu}(t)$. To define
them, following \cite{ya1,ya2,ya3}, we rewrite
Eqs.\ref{ee}-\ref{qf} in the form
\begin{equation}
\begin{array}{c}
\left\langle \exp\,\int _0^t \left[v_{\tau }^{\mu \nu } x_{\mu \nu
}(\tau )+f_{\tau }^{\mu \nu }y_{\mu \nu }(\tau )\right]
d\tau\right\rangle_o\equiv \\
\equiv \Xi (V_{\tau };\,H_0,F_{\tau })=\\
=\text{Tr\,}\,\,\overleftarrow{\exp } \left(\int _0^t\left[ -\frac
i\hbar (H_0-F_{\tau })+\frac 12 V_{\tau }\right] d\tau
\right)\,\rho _0\times \\
\,\,\,\,\,\,\,\,\,\times \,\overrightarrow{\exp } \left(\int
_0^t\left[ \frac i\hbar (H_0-F_{\tau })+\frac 12 V_{\tau }\right]
d\tau \right)
\end{array}  \label{qfxy}
\end{equation}
Essentially, it is assumed here that an imaginative probability
measure hidden behind the angle brackets is itself independent of
the forces. In other words, all the random processes
$\,x_{\mu\nu}(t)\,$ and $\,y_{\mu\nu}(t)\,$ are meant be
characteristics of the unperturbed dynamics governed by the
Hamiltonian $\,H_0$. The subscript ``o'' serves to remind of this
circumstance.

Therefore, it may be convenient to rewrite (\ref{qfxy}) once more,
in the form
\begin{equation}
\begin{array}{c}
\left\langle \exp\,\int _0^t \left[v_{\tau }^{\mu \nu } x_{\mu \nu
}(\tau )+f_{\tau }^{\mu \nu }y_{\mu \nu }(\tau )\right]
d\tau\right\rangle_o=\\
=\Xi (V_{\tau };\,H_0,F_{\tau })=\\
=\text{Tr\,}\,\,\overleftarrow{\exp } \left(\int _0^t\left[ \frac
i\hbar f_{\tau }^{\mu\nu}+\frac 12 v_{\tau
}^{\mu\nu}\right]X_{\mu\nu}^o(\tau) d\tau
\right)\,\rho _0\times \\
\,\,\,\,\,\,\,\,\,\times \,\overrightarrow{\exp } \left(\int
_0^t\left[-\frac i\hbar f_{\tau }^{\mu\nu}+\frac 12 v_{\tau
}^{\mu\nu}\right]X_{\mu\nu}^o(\tau) d\tau \right)\,\,,
\end{array}  \label{fxy}
\end{equation}
where $\,X_{\mu\nu}^o(t)\,$ are the operators $\,X_{\mu\nu}\,$
considered in the interaction representation, thus representing
free evolution:
\[
X_{\mu\nu}^o(t)=U_0^{-1}(t)X_{\mu\nu}U_0(t)= X_{\mu\nu}\exp\,
(iE_{\mu\nu}t/\hbar )
\]

According to (\ref{qfxy})-(\ref{fxy}), if $\,V_t\equiv 0\,$ then
\begin{equation}
\begin{array}{c}
\left\langle \exp\,\int _0^t f_{\tau }^{\mu \nu }y_{\mu \nu }(\tau
)\,d\tau\right\rangle_o=\Xi (0;H_0,F_{\tau })=1\,\,,
\end{array}  \label{ymoms}
\end{equation}
that is any statistical moment of $\,y$'s themselves is equal to
zero. But their correlations with $\,x$'s differ from zero:
\begin{equation}
\left\langle \prod _{j,m}x(t_j)y(\tau _m)\right\rangle_o = \left [
\prod _m\frac {\delta }{\delta f(\tau _m)} \left\langle \prod_j
x(t_j)\right\rangle_{F_{\tau}} \right ] _{F_{\tau}=0}
\label{cors0}
\end{equation}
(indices $\,\mu ,\nu\,$ are omitted). Hence, cross-correlation
between $\,N\,$ copies of $\,x\,$ and $\,M\,$ copies of $\,y\,$
represents $\,M$-order response to perturbations of an $\,N$-order
statistical moment of the $\,x$'s.

Interestingly, the relations (\ref{cors0}) are valid also for the
$\,x$'s and $\,y$'s cumulants (semiinvariants) whose generating
function is $\,\ln\,\Xi (V_{\tau };H_0,F_{\tau })$. The proof is
trivial:
\[
\begin{array}{c}
\ln\,\left\langle \exp\,\int \left[v_{\tau }x(\tau )+
f_{\tau }y(\tau )\right]\,d\tau \right\rangle _o=\\
=\left[\exp \left (\int d\tau f_{\tau }\,\delta/\delta
g_{\tau}\right )\, \ln\,\Xi (V_{\tau };\,H_0,G_{\tau
})\right]_{G_{\tau }=0}
\end{array}
\]
(again without indices). Consequently, instead of (\ref{cors0}) we
can write
\begin{equation}
\left\langle \prod _{j,m}x(t_j)y(\tau _m)\right\rangle_o^c = \left
[ \prod _m\frac {\delta }{\delta f(\tau _m)} \left\langle \prod_j
x(t_j)\right\rangle_{F_{\tau}}^c \right ] _{F_{\tau}=0}
\label{cors}
\end{equation}
with the superscript ``c'' marking the cumulants.

The union of the two sets of random processes, quite realistic
$\,x$'s and indeed rather illusive $\,y$'s, gives stochastic
representation of the system's response to perturbations. If the
latter are caused by interactions with some other dynamical system
``D'', then we make first step towards the stochastic
representation of deterministic (quantum or classical)
interactions, which was suggested in \cite{ya1} and developed in
\cite{ya2,ya3,ya4}. For instance, our system can serve as
thermostat for ``D''. General stochastic equations which describe
``D'' under influence by the thermostat inevitably include the
$\,y$'s whose main effect is dissipation.

\section{Time reversal and generating FDR in the stochastic representation}
Combining Eqs.\ref{fds}-\ref{nv1} with Eq.\ref{qfxy}, one can
simply reformulate the generating FDR in terms of $\,x$'s and
$\,y$'s~:
\begin{equation}
\begin{array}{c}
\left\langle \exp\,\int _0^t \left[v_{\tau }^{\mu \nu }x_{\mu \nu
}^{\prime}(t-\tau )+f_{\tau }^{\mu \nu }y_{\mu \nu
}^{\prime}(t-\tau)\right]d\tau\right\rangle_o=\\
=\left\langle \exp\,\int _0^t \left[v_{\tau }^{\mu \nu }
\widetilde{x}_{\mu \nu }(\tau )+f_{\tau }^{\mu
\nu}\widetilde{y}_{\mu \nu }(\tau
)\right]d\tau\right\rangle_o\,\,,
\end{array}  \label{fdsxy}
\end{equation}
where, of course, $\,x_{\mu \nu }^{\prime}=x_{\nu \mu }\,$,
$\,y_{\mu \nu }^{\prime}=y_{\nu \mu }$, and
$\,\widetilde{x},\,\widetilde{y}\,$ relate to $\,x,\,y\,$
absolutely similar to (\ref{nv2}):
\begin{equation}
\left[
\begin{array}{c}
\widetilde{x}_{\mu \nu}(\tau) \\
\widetilde{y}_{\mu \nu}(\tau)
\end{array}
\right]= \left[
\begin{array}{cc}
C_{\mu \nu } & \frac {i\hbar }{2} S_{\mu \nu } \\
\frac {2}{i\hbar }S_{\mu \nu } & C_{\mu \nu }
\end{array} \right] \left[
\begin{array}{c}
x_{\mu \nu}(\tau) \\ y_{\mu \nu}(\tau)
\end{array} \right]      \label{nvxy}
\end{equation}
Evidently, it the matrices $\,x=\{x_{\mu\nu}\}\,$ and
$\,y=\{y_{\mu\nu}\}\,$ are Hermitian, $\,x^{\dagger }=x\,$ and
$\,y^{\dagger}=y\,$, then the transformation (\ref{nvxy}) does not
damage this their property.

Taking into account $\,\rho_0$'s invariance  with respect to free
evolution, the FDR (\ref{fdsxy}) can be expressed also in the form
\begin{equation}
\begin{array}{c}
x_{\nu \mu }(t_0-t)\asymp \widetilde{x}_{\mu
\nu}(t)\,\,,\,\,\,y_{\nu \mu }(t_0-t)\asymp \widetilde{y}_{\mu
\nu}(t)\,\,,
\end{array}       \label{stocheq}
\end{equation}
where $\,t_0\,$ is arbitrary time shift, and symbol $\,\asymp \,$
means statistical equivalence of left- and right-handed random
processes (i.e. equivalence in the sense of statistical moments,
$\,\langle x^{\prime}(t_0-t_1)...x^{\prime}(t_0-t_N)\rangle$
$=\langle \widetilde{x}(t_1)...\widetilde{x}(t_N)\rangle\,$, and
so on).

Following \cite{ya1}, it may be convenient to introduce another
random processes, whose matrices are non-Hermitian:
\begin{equation}
\begin{array}{c}
\xi(t)\equiv x(t)+\frac {i\hbar}{2}y(t)\,\,,\\\,\\ \eta(t)\equiv
\xi ^{\dagger}(t)=x(t)-\frac {i\hbar}{2}y(t)\,
\end{array} \label{xieta}
\end{equation}
In their terms the generating FDR look most simple:
\begin{equation}
\begin{array}{c}
\xi_{\nu \mu }(t_0-t)\,\asymp\,\exp (
E_{\mu\nu}/2T)\,\xi_{\mu\nu}(t)\,\,\,,
\end{array}       \label{stocheq1}
\end{equation}
\begin{equation}
\begin{array}{c}
\eta_{\nu \mu }(t_0-t)\,\asymp\,\exp
(-E_{\mu\nu}/2T)\,\eta_{\mu\nu}(t)
\end{array}       \label{stocheq2}
\end{equation}

To some extent, the $\,\xi$'s and $\,\eta$'s can be thought like
amplitudes of quantum jumps, hence, their squares, $\,|\xi|^2$ and
$\,|\eta|^2$, like corresponding probabilities. Then, formulas
(\ref{stocheq1}) and (\ref{stocheq2}) reduce to familiar relations
between probabilities of mutually reversed jumps.

It is necessary to remember, of course, that generally (when
$\,H_0^{\prime}\neq H_0\,$) the left and right-hand processes in
(\ref{stocheq}), (\ref{stocheq1}) and (\ref{stocheq2}) relate to
somehow different systems.

Combining these statistical equalities with (\ref{cors0}) (or
(\ref{cors})) and (\ref{qfxy}) (or (\ref{fxy})), and besides with
the causality principle, as well as with independence of
statistical moments on $\,t_0\,$, one can construct relatively
simple algorithms for derivation of many particular FDR. However,
that are tasks for separate work.

\section{Conclusion}
To resume, we obtained generating fluctuation-dissipation
relations (FDR) for a dynamically perturbed quantum system,
assuming special but theoretically and practically important
definition of its quantum statistical moments (or corresponding
characteristic functional) which describe consecutive or
continuous measurements of the system.

In addition, short and expressive formulation of the FDR in terms
of the stochastic representation of quantum interactions was done.





\end{document}